\documentclass[aps, pre, reprint, groupedaddress, showpacs, twocolumn]{revtex4-1}

\usepackage{amsmath,amssymb,bm}
\usepackage{graphicx}
\usepackage{color}
\usepackage{calligra}
\usepackage{ulem}

\usepackage[dvipdfmx,
colorlinks=true,
urlcolor=blue,
citecolor=blue,
linkcolor=blue,
hyperfootnotes=false]{hyperref}

\begin{document}

\title{Finite-$m$ scaling analysis of Berezinskii-Kosterlitz-Thouless phase transitions and entanglement spectrum for the six-state clock model}

\author{Hiroshi Ueda$^{1,2}$}
\author{Kouichi Okunishi$^3$}
\author{Kenji Harada$^4$}
\author{Roman Kr\v{c}m\'ar$^5$}
\author{Andrej Gendiar$^5$}
\author{Seiji Yunoki$^{1,6,7}$}
\author{Tomotoshi Nishino$^{8}$}

\affiliation{$^1$Computational Materials Science Research Team, 
RIKEN Center for Computational Science (R-CCS), Kobe 650-0047, Japan}
\affiliation{$^2$JST, PRESTO, Kawaguchi, 332-0012, Japan}
\affiliation{$^3$Department of Physics, Niigata University, Niigata 950-2181, Japan}
\affiliation{$^4$Graduate School of Informatics, Kyoto University, Kyoto 606-8501, Japan}
\affiliation{$^5$Institute of Physics, Slovak Academy of Sciences, SK-845 11, Bratislava, Slovakia}
\affiliation{$^6$Computational Condensed Matter Physics Laboratory, RIKEN Cluster for Pioneering Research (CPR), Wako, Saitama 351-0198, Japan}
\affiliation{$^7$Computational Quantum Matter Research Team, 
RIKEN Center for Emergent Matter Science (CEMS), Wako, Saitama 351-0198, Japan}
\affiliation{$^8$Department of Physics, Graduate School of Science, Kobe University, Kobe 657-8501, Japan}

\date{\today}

\begin{abstract}
We investigate the Berezinskii-Kosterlitz-Thouless transitions for the square-lattice six-state clock model with the corner-transfer matrix renormalization group (CTMRG).
Scaling analyses for effective correlation length, magnetization, and entanglement entropy with respect to the cutoff dimension $m$ at the fixed point of CTMRG provide transition temperatures consistent with a variety of recent numerical studies. 
We also reveal that the fixed point spectrum of the corner transfer matrix in the critical intermediate phase of the six-state clock model is characterized by the scaling dimension consistent with the $c=1$ boundary conformal field theory associated with the effective $Z_6$ dual sine-Gordon model.
\end{abstract}

%\pacs{05.10.Cc, 02.70.-c, 05.30.-d, 71.27.+a}

\maketitle

\section{Introduction}
The two dimensional (2D) $q$-state clock models have been providing interesting phase-transition physics competingly induced by classical orders due to $q$-sided polygon-type discretization and the Berezinskii-Kosterlitz-Thouless (BKT) phase associated with the classical XY model~\cite{berezinskii, kosterlitz}.
For $q=2$ and 3, the second-order transitions of Ising universality and three-state Potts universality are respectively confirmed. 
The $q=4$ clock model is equivalent to the two decoupled Ising models.
For $q \geq 5$, meanwhile, the critical intermediate phase is expected between the ordered and disordered phases, accompanying the BKT transitions at the boundaries of the intermediate phase~{\cite{eta_for_z6,tobochnik,challa,yamagata,tomita,brito,baek1,baek2,kumano,Kromar2016,chen_p6,Surungan2019,Hong2019,Li2019}.
However, some studies of Monte Carlo (MC) simulations are controversial~\cite{Lapilli2006,hwang}. 
This is basically because the BKT transitions exhibit very weak singularity near the transition points. 
Thus, numerical simulations for finite size systems often suffer from the logarithmic dependence in the finite-size-scaling (FSS) analyses of bulk physical quantities such as specific heat and  order parameters.
Thus, precise verification of the BKT transitions for the clock models has been a challenging problem in the context of computational physics.

In this paper, we focus on the critical intermediate phase and the BKT transitions of the six-state clock model, for which a variety of numerical investigations were performed. 
The current status of numerical estimations of the transition points is summarized in Table \ref{Table:Tc_List}.
MC simulations combined with the FSS analysis for the correlation length, a ratio of the spin-spin correlation function, helicity modulus, and roughness of the spin~\cite{Brito2007} provide a couple of estimations for the BKT transitions: the lower and upper transition temperatures are respectively located at  $T_{c1} \sim 0.7$ and  $T_ {c2} \sim 0.9$.
However, larger system sizes are still required for the precise determination of the transition temperatures. 
Recently, tensor network approaches were also tested in cooperation with the scaling analysis for the entanglement entropy~\cite{Kromar2016, chen_p6, Li2019}  and Fisher zero~\cite{Hong2019}. However, some results for $T_{c2}$ seems to slightly deviate from the recent MC results. 
Moreover, the BKT-transition nature also makes it difficult to numerically check the scaling dimensions associated with $c=1$ conformal field theory (CFT), which is the effective field theory describing the critical intermediate phase.

\begin{table*}[hbt]
  \caption{List of the lower and upper transition temperatures, $T_{\rm c1}$ and $T_{\rm c2}$.}
\label{Table:Tc_List}
\begin{tabular}{lllll}
\hline\hline
&  Method & $L$ or $m$ &  $T_{c1}$     &  $T_{c2}$  \\
\hline
Tobochnik~\cite{tobochnik}(1982)  & MCRG  & $L=32$ &     0.6      &    1.3        \\
Challa and Landau~\cite{challa}(1986)        &MC & $L=72$ &     0.68(2)   &    0.92(1)    \\
Yamagata and Ono~\cite{yamagata}(1991)  &MC &   &     0.68      &    0.90       \\
Tomita and Okabe~\cite{tomita}(2002)       &Probability-changing cluster MC & $L=512$  &    0.7014(11)  &    0.9008(6)  \\
Hwang~\cite{hwang}(2009)                      &Wang-Landau MC& $L=28$  &     0.632(2)   &    0.997(2)   \\
Brito {\it et al.}~\cite{brito}(2010)          &Heat-bath single spin flipping MC& $L=160$ &  0.68(1)     &    0.90(1)    \\
Baek {\it et al.}~\cite{baek1,baek2}(2010)    &Wolff MC&  $L=512$  &       -         &    0.9020(5)  \\
Kumano {\it et al.}~\cite{kumano}(2013)       &Boundary-flip MC& $L=32(256)$   &     0.700(4)    &    0.904(5)   \\
Kr\v cm\'ar {\it et al.}~\cite{Kromar2016}(2016)  &CTMRG& $L=129$ &     0.70        &    0.88   \\
Chen {\it et al.}~\cite{chen_p6}(2017)  &HOTRG& $m=15$ &     0.6658(5)  &    0.8804(2) \\
Surungan {\it et al.}~\cite{Surungan2019}(2019)  &Swendsen-Wang MC & $L=512$ &     0.701(5)        &    0.898(5)   \\
Hong and Kim~\cite{Hong2019}(2019)  &HOTRG & $L=128$ &     0.693        &    0.904   \\
Li {\it et al.}~\cite{Li2019}(2019)  &VUMPS & $m=250$ &     0.6901(4)        &    0.9127(5)   \\
\hline
This work   & CTMRG (Correlation length etc.) & $m=768$ &  0.694(3)   & 0.908(3) \\
  & CTMRG (Entanglement Spectrum)& $m=768$ &  0.693   & 0.900 \\
\hline\hline
\end{tabular}
\end{table*}

For the six-state clock model, we therefore perform large-scale corner-transfer-matrix-renormalization-group (CTMRG) calculations~\cite{ctmrg1,ctmrg2} up to the cutoff dimension $m=768$, with use of a parallelized solver of the matrix-eigenvalue problem~\cite{pctmrg}.
In particular, we employ the finite-$m$ scaling analysis for the fixed-point of CTMRG calculations with various $m$, which has been successfully applied to second-order transitions~\cite{NOK,tagliacozzo,HU2017}. 
For the present case,  based on $m$-dependence of the effective correlation length, we practically estimate transition temperatures and scaling dimensions for magnetization and ``classical analogue of entanglement entropy'' simply referred to as ``entanglement entropy'' hereafter, with assuming the scaling form of the BKT transition.
The estimated transition temperatures are listed in Table~\ref{Table:Tc_List}, which are basically consistent with the recent MC and tensor network results.
We also address the scaling analysis for the entanglement spectra determined by the corner-transfer-matrix (CTM) in the intermediate phase, assuming the boundary CFT with the $c=1$ Gaussian universality.
We then show that the Tomonaga-Luttinger (TL) parameter extracted from the entanglement spectra is consistent with the result of the effective $Z_6$ dual Sine-Gordon field model for the six-state clock model~\cite{Matsuo2006,Li2019}.

The organization of the rest of this paper is as follows.
In the next section, we explain the setup of CTMRG and present numerical results for the correlation length, entanglement entropy, and  magnetization. 
In Sec.~\ref{sec:fms}, we explain a BKT version of the finite-$m$ scaling.  
We also show the results of finite-$m$ scaling analysis with the phenomenological renormalization group (PRG)~\cite{Hida} to reduce subleading effects.
In Sec.~\ref{sec:es}, we show the scaling analysis for the entanglement spectrum based on the boundary CFT and discuss the consistency between the numerical results with the effective field theory.
In Sec.~\ref{sec:summary}, a summary and prospects of this study are presented.

\section{Numerical calculation}

\subsection{CTMRG}

In this work, we use CTMRG to calculate the spontaneous magnetization, the correlation length, and the entanglement entropy for the six-state clock model on a square lattice. 
We write the local Hamiltonian for the nearest neighboring sites as
\begin{equation}
	{\cal H}_{ab} = - J \cos\left( \frac{2\pi}{q}(a - b) \right) \, ,
\end{equation}
with $q = 6$, where $J$ denotes the exchange coupling and the indices $a$ and $b$ $= 1, 2, \cdots 6 $ specify clock angles. In the following, we assume $J=1$ for simplicity.
Then, the local Boltzmann weight of the six-state clock model is practically represented as 
\begin{equation}
W_{abcd}^{~} = G_{ab}^{~} G_{bc}^{~} G_{cd}^{~} G_{da}^{~} \, ,
\label{eq:W}
\end{equation}
which can be regarded as a local 4-leg vertex tensor on the 45$^\circ$ rotated square lattice. 
Here, the bond weights $G_{ab}$ is defined with
\begin{equation}
G_{ab}^{~} = \exp \left( -\beta {\cal H}_{ab} \right),
\end{equation}
where $\beta\equiv 1/T$ is the inverse temperature. Note that we have assumed $k^{~}_{\rm B}=1$.

In the CTM formulation, the partition function of the system is represented as 
$
Z \equiv \mathrm{Tr} \, C^4,
$
where $C$ denotes the renormalized CTM.
Since Eq. (\ref{eq:W}) has the $\pi/2$ rotation and parity symmetries, the CTMs corresponding to the four quadrants of the lattice are equivalent.
In CTMRG, then, we recursively update the CTM and the half row-to-row (column-to-column) transfer matrices toward the bulk fixed point, using the transformation matrix provided with diagonalization of the CTM.
Here, note that free or ferromagnetic boundary conditions can be appropriately set up with initial transfer matrices. 
After a sufficient number of iterations, we obtain the fixed point matrices and then evaluate the bulk magnetization 
$
M(T,m)
$ 
and the entanglement entropy
$
S_{\rm E}^{~}(T,m) = - {\rm Tr} \, \rho  \ln \, \rho 
$
with $\rho \equiv C^4/Z$.
A typical number of iterations for the convergence is of the order of $10^4$ near the transition points for $m=768$. 
The numerical accuracy of $M(T,m)$ and $S^{~}_{\rm E}(T,m)$ at the fixed point is, of course, governed by the cutoff dimension $m$.
The truncation error due to the cutoff $m$ is basically equivalent to that of tensor-network algorithms based on the matrix product state.

In order to evaluate the typical length scale of the fixed point with a finite $m$, moreover, we can extract the effective correlation length as 
\begin{equation}
\xi_{\rm e}( T, m ) = \Bigl[ \ln\bigl( \zeta_1^{~} / \zeta_2^{~} \bigr) \Bigr]^{-1}_{~} \, ,
\label{eq:correlation_length}
\end{equation}
where $\zeta_1^{~}$ and $\zeta_2^{~}$ respectively denote the largest- and second-largest eigenvalues of the renormalized row-to-row transfer matrix constructed from the renormalized half-row transfer matrix at the fixed point.
Note that $\xi_{\rm e}(T,m)$ has a finite value even in the critical phase since the finite-$m$ effect gives rise to  an effective length scale. 
In the next section, this effective correlation length $\xi_{\rm e}$ plays an essential role in performing the finite-$m$ scaling analysis for $M(T,m)$ and $S^{~}_{\rm E}(T,m)$.

\subsection{results}

In Fig.~\ref{fig:mag_vs_t}, we first present temperature dependences of the magnetization $M(T,m)$ for the ferromagnetic boundary condition.
In the low-temperature region ($T\lesssim 0.7$), the fact that $M(T,m)$ has no $m$ dependence gives clear evidence of the spontaneous breaking of the $Z_6$ symmetry. In contrast, the high-temperature region ($ T \gtrsim 1.0$) is in the disordered phase.
Moreover, we find that the shoulder structure of $M(T,m)$ accompanying the strong $m$ dependence appears between $0.7 \lesssim T \lesssim 1.0$, suggesting that the critical intermediate phase is actually the case for the six-state clock model.
This is because the finite-$m$ effect and the symmetry breaking boundary condition may induce a finite $M(T,m)$ even in the critical regime. 
Note that in Fig.~\ref{fig:mag_vs_t}, we also show the lower and upper transition temperatures $T_{c1}=0.694$ and $T_{c2}=0.908$ as the vertical lines in advance, which will be estimated with the scaling analysis in the next section.

\begin{figure}
\includegraphics[width=8cm]{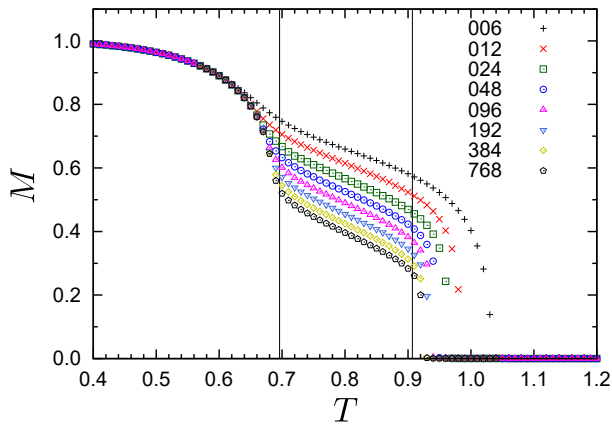}
\caption{(Color online) Temperature dependence of the magnetization $M(T,m)$ for $m=6,12,\cdots,768$ under the ferromagnetic boundary condition. The vertical lines indicate the lower and upper transition points estimated with the finite-$m$ scaling analysis.}
\label{fig:mag_vs_t}
\end{figure}

\begin{figure}
\includegraphics[width=8cm]{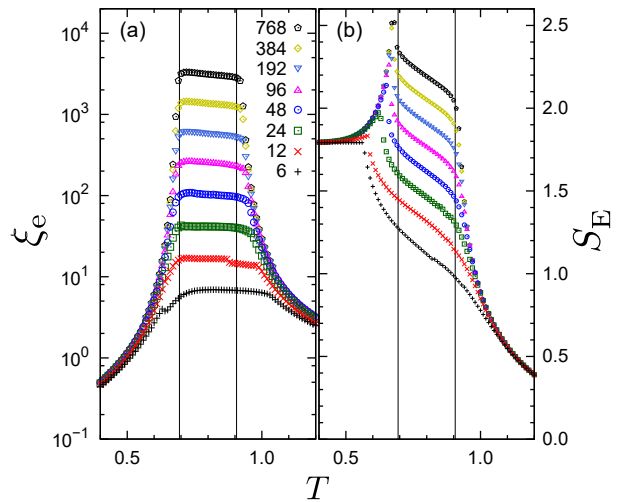}
\caption{(Color online) (a) Temperature dependence of the correlation length $\xi_{\rm e}(T,m)$ for the ferromagnetic boundary
(b) the entanglement entropy $S^{~}_{\rm E}(T,m)$ for the free boundary.
The curves with symbols represent the CTMRG results with cutoff dimensions $m=6,12,\cdots,768$ from bottom to top.
}
\label{fig:xi_ee_vs_t}
\end{figure}

In order to analyze the critical intermediate phase, behaviors of the correlation length are essential.
In Fig.~\ref{fig:xi_ee_vs_t}(a), we next show the correlation length $\xi_{\rm e}(T,m)$ for the ferromagnetic boundary condition, where $\xi_{\rm e}(T,m)$ exhibits the plateau like-behavior in the intermediate region. 
As $m$ increases, $\xi_{\rm e}$ increases with a power-law behavior, which also suggests the critical intermediate phase consistent with the magnetization result.
In Fig.~\ref{fig:xi_ee_vs_t}(b), we finally show the entanglement entropy $S^{~}_{\rm E}(T,m)$ for the free boundary condition.
For the free-boundary case, the fixed-point CTM in the ordered phase equivalently includes the contributions from the six broken-symmetry states, implying that the eigenvalue spectrum of CTM has the six-fold degeneracy.
Thus, $S^{~}_{\rm E}(T,m)$ in the low-temperature limit is $S^{~}_{\rm E}= \log 6 =1.79$. 
In the intermediate temperature region,   $S_{\rm E}^{~}$ also exhibits a diverging behavior with respect to $m$.
According to a CFT,  the bipartition entanglement entropy in the critical regime can be described as  
\begin{equation}
S_{\rm E}^{~} \simeq \frac{c}{6} \, \ln \, \ell + {\rm const.}\,,
\label{eq:SE_scaling}
\end{equation}
where $c$ denotes the central charge, and $\ell$ is the length of the system part~\cite{Vidal, Calabrese}.
Replacing $\ell \to \xi_{\rm e}(T,m)$ in Eq. (\ref{eq:SE_scaling}), then, we can basically understand  behavior of $S^{~}_{\rm E}(T,m)$ in Fig. \ref{fig:xi_ee_vs_t}(b),  
except for the contribution from the nonuniversal constant.
Nevertheless,  a finite-$m$ scaling analysis is required for extracting  precise critical behaviors in the intermediate region and the BKT phase boundaries in Figs. \ref{fig:mag_vs_t} and \ref{fig:xi_ee_vs_t}.

\section{Finite-$m$ scaling  and BKT phase transitions}
\label{sec:fms}
For the BKT transition,  the correlation length diverges with $\xi \sim \exp ( {\rm const.} \times t^{-1/2}) $ toward the transition point $T_c$, where $t \equiv |T/T_c - 1|$ denotes the normalized temperature.
However, the conventional scaling hypothesis based on the divergence of $\xi$ often encounters difficulty in a precise determination of the transition point, since the essential singularity of $\xi$ induces very weak anomaly in bulk observables such as specific heat and magnetization.

In order to perform a stable scaling analysis for CTMRG data with finite $m$,  we phenomenologically assume the scaling form for a certain quantity $A$ ($A \in M$ or $S^{~}_{\rm E}$) with the scaling dimension $x_A$ as
\begin{equation}
A( t,\ell ) = \ell^{x_A} \, f_A^{~} \left( t \left(\ln \frac{\ell}{\epsilon} \right)^2 \right) \,
\label{eq:scaling}
\end{equation}
where $\ell$ denotes a characteristic length of the system particularly in numerical simulations,  $\epsilon $ is a cutoff scale, and $f_A( y )$ is a scaling function.
This scaling form of Eq. (\ref{eq:scaling}) was originally proposed for the Helicity modulus of the 2D XY model  in Ref.~[\onlinecite{Harada-Kawashima}], based on the renormalization group (RG) flow for the effective Sine-Gordon model, and later was applied to the magnetization of the clock models in Ref.~[\onlinecite{chatelain}].

In the critical phase,  the effective length scale at the CTMRG fixed point with a cutoff dimension $m$ is given by  Eq.~(\ref{eq:correlation_length}). 
Then, an essential point is that the asymptotic behavior of $\xi_{\rm e}$ with respect to $m$ is also described by the power-law~\cite{tagliacozzo},  
\begin{align}
\xi_{\rm e} \sim  m^\kappa,
\label{eq:xi_kappa}
\end{align}
where  $\kappa$ denotes an exponent characteristic to the matrix-product-state description of the eigenvector of the row-to-row transfer matrix. 
We can perform fitting of Eq. (\ref{eq:xi_kappa}) for the CTMRG results of $\xi_{\rm e}$ in Fig. \ref{fig:xi_ee_vs_t}, using data in $m =48 \sim 768$.
We then extrapolate the fitting result to $m\to \infty$ based on the phenomenological renormalization group (PRG)~\cite{Hida} and obtain $\kappa \simeq 1.17$ in the critical phase [Fig.~\ref{fig:kappa}].
An important point for this result is that the value of $\kappa$ is basically invariant in the intermediate critical phase, which is often observed for $c=1$ CFTs.
Indeed, $\kappa \sim 1.16$ and $c \sim 0.985$ were reported for such a quantum spin system as chiral ladder in the critical regime~\cite{schmoll_orus_prb_2019}, which is consistent with the present result for the clock model.
This result is also supported by a general discussion based on the matrix product state~\cite{pollmann} implying that $\kappa$ depends only on $c$ through 
\begin{equation}
\kappa = \frac{6}{ c\bigl( \sqrt{12 / c} \, + 1 \bigr) }\, ,
\label{pollmann}
\end{equation}
although $\kappa \simeq 1.17$ slightly deviates from the value of Eq. (\ref{pollmann}) for $c=1$.

\begin{figure}
\includegraphics[width=7cm]{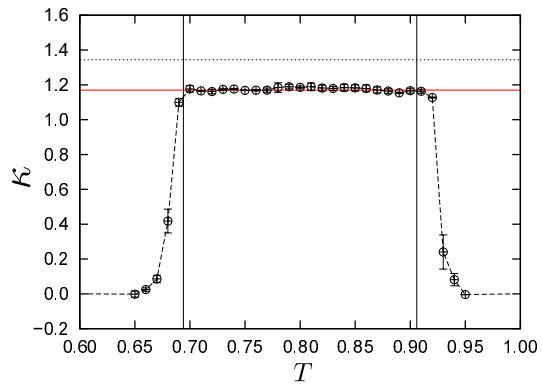}
\caption{(Color online) Critical exponent $\kappa$ extracted with finite-$m$ scalings for the effective correlation length $\xi_{\rm e}\sim m^\kappa$.}
\label{fig:kappa}
\end{figure}

Taking account of the divergence of the effective correlation length, we may substitute $\ell \sim  \xi_{\rm e}\sim m^\kappa$ into Eq.~(\ref{eq:scaling}).
For the magnetization, we then write the scaling ansatz in the vicinity of the BKT transition point as 
\begin{eqnarray}
M( t, m) & = & m^{- \kappa \eta/2 }_{~} \tilde{f}_M^{~}\left( t \left(\ln \frac{m}{\epsilon}  \right )^2 \right),
\label{eq:scaling_m}
\end{eqnarray}
where  $\tilde{f}_M^{~}$ is a scaling function satisfying $\tilde{f}_M^{~}( y ) \sim {\rm const.}$ for $y \ll 1$, and  the cutoff scale $\epsilon$ was redefined.
Note that $\eta(=-2 x_M)$ is the anomalous dimension, which is consistent with the scaling relations used in Refs. [\onlinecite{challa,Borisenko,chatelain}].

The entanglement entropy is not a directly observable quantity.
However,  Eq.~(\ref{eq:SE_scaling}) suggests that a scaling dimension of $ e^{S_{\rm E}^{~}} $ may be regarded as $c/6$.
Thus, plugging $\ell \sim m^\kappa$ into  Eq. (\ref{eq:scaling}) with $x_A=c/6$, we assume the scaling form  for $e^{S_{\rm E}^{~}}$ as
\begin{eqnarray}
e^{S_{\rm E}(t,m)}_{~} \sim   m^{c \kappa / 6}_{~} \, g\left( t \left(\ln \frac{m}{\epsilon} \right)^2_{~} \right) \, ,
\label{eq:scaling_ee}
\end{eqnarray}
where ${g}$ is a scaling function.

\begin{figure}
\includegraphics[width=8cm]{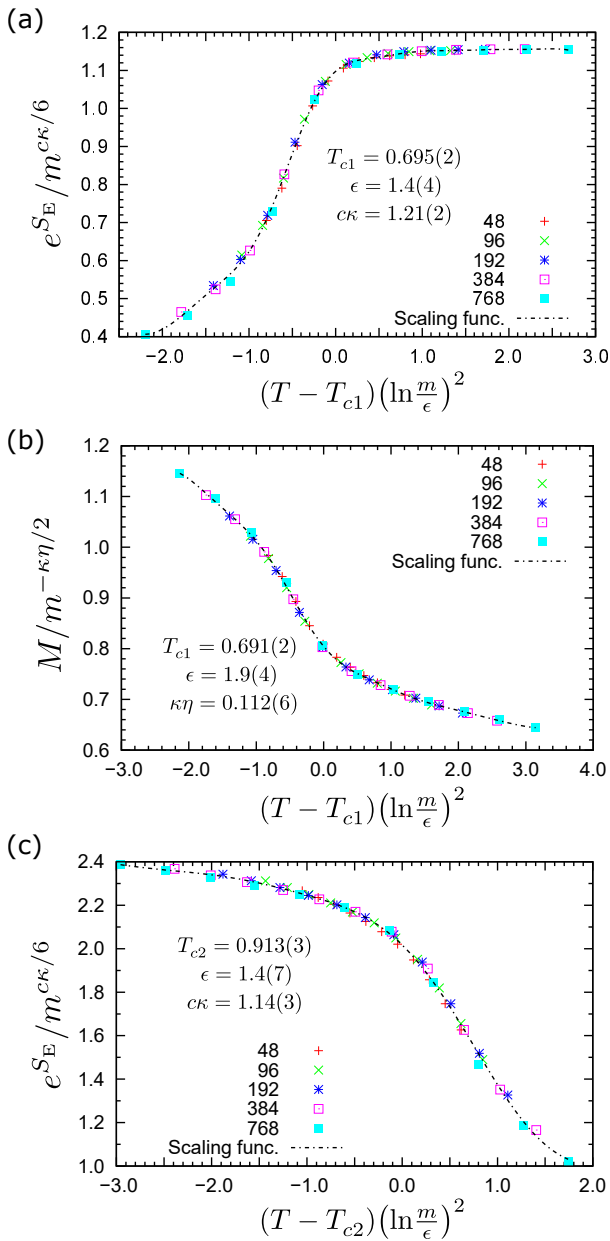}
\caption{(Color online) Finite-$m$ scaling plots for the BKT transition points of the six-state clock model:
(a) Entanglement entropy and (b) magnetization for $T_{c1}$, and (c) entanglement entropy for $T_{c2}$. }
\label{fig:scaling}
\end{figure}

Using the scaling forms of Eqs. (\ref{eq:scaling_m}) and (\ref{eq:scaling_ee}), we perform finite-$m$ scalings with the use of the Bayesian inference algorithm~\cite{Harada}, respectively  for $M$ and $e^{S^{~}_{\rm E}} $ of $m =46$, 96, 192, 384, and 768.
Note that for $T_{c1}$,  $M$ and $S_{\rm E}$ with the ferromagnetic boundary condition are used, and the temperature range for the fitting is $T=[0.65,0.75]$. 
Meanwhile, for  $T_{c2}$,  those with the free boundary condition are used with the fitting window $T=[0.85,0.95]$.
The results of the finite-$m$ scaling plots are shown in Fig. \ref{fig:scaling}, where the data for various $m$ are basically collapsed on scaling functions.
The estimated transition points are $T_{c1}=0.695(2)$ and $T_{c2}=0.913(3)$ for the entanglement entropy and $T_{c1}=0.691(2)$ for the magnetization.
Also we obtain $c\kappa = 1.21(2)$ and $\kappa\eta =0.112(6)$ for $T_{c1}$, and   $c\kappa = 1.14(3)$ for $T_{c2}$.
However, we should note that these transition points and exponents still contain weak $m$ dependencies, which suggest that  corrections to Eqs.~(\ref{eq:scaling_m}) and (\ref{eq:scaling_ee}) may not be negligible.

In order to extrapolate the transition points and exponents in the $m\to \infty$ limit,  we employ the PRG~\cite{Hida}. 
In the PRG, we firstly estimate $\mathcal{O} \in \{ T_{\rm c1}, c\kappa, \eta \kappa, T_{\rm c2} \}$ with Eqs. (\ref{eq:scaling_m}) and (\ref{eq:scaling_ee}) for $m_1$ and $m_2( \neq m_1)$ and interpolate $\mathcal{O}^{*}$ at $m^{*}=(m_1+m_2)/2$.
We nextly plot $\mathcal{O}^{*}$ as functions of $1/m^{*}$ and then perform the polynomial fitting for $\mathcal{O}^{*}$ with respect to $1/m^{*}$ to extrapolate  $\lim_{m^{*} \rightarrow \infty} \mathcal{O}^{*}$. 
Figure~\ref{fig:prg} shows $ T^{*}$ as functions of $1/m^{*}$. 
In Fig. \ref{fig:prg}(a), for example,  $m^{*}$ dependencies of $T^{*}_{\rm c1}$ for $M$ and $e^{S_{\rm E}}$ tend to converge in the limit of $m^{*} \rightarrow \infty$. 
Thus, we extrapolate $T^{*}_{\rm c1}$ for $1/m^{*} <0.007$ including the upper and lower boundaries of the error bars, which are depicted as guidelines in the figure,  and obtain  $T_{\rm c1}=0.694(3)$ at $m^{*}\to \infty$. 
For the upper transition point, the similar analysis also yields $T_{\rm c2}=0.908(3)$.
These transition points of $T_{\rm c1}$ and $T_{\rm c2}$ are consistent with the results of recent works listed in Table~\ref{Table:Tc_List}.
Moreover, we also obtain $c=0.97(3)$ and $\eta=0.09(1)$ for $M$ at $T_{\rm c1}$, which are basically consistent with the theoretical values, $c=1$ and $\eta=1/9=0.11\cdots$~\cite{eta_for_z6,Matsuo2006}.

\begin{figure}
\includegraphics[width=8cm]{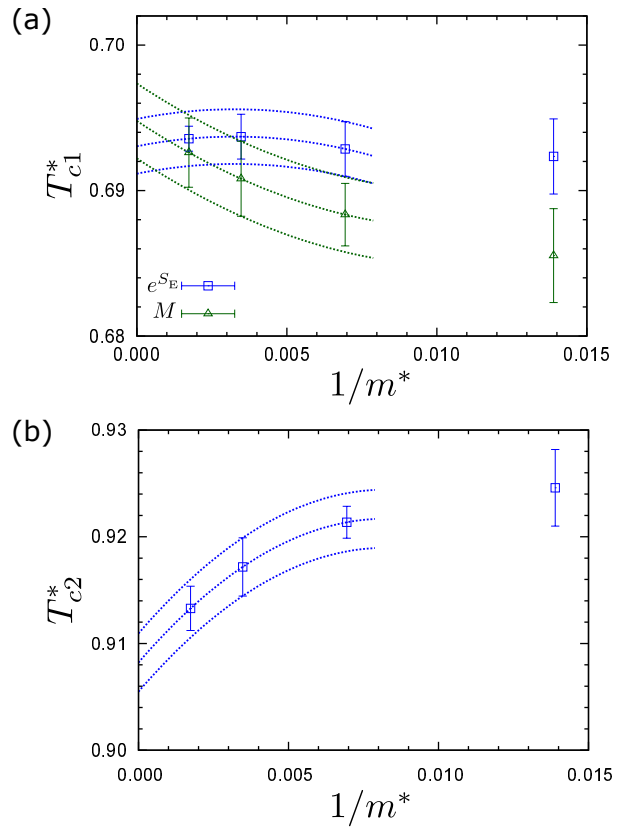}
\caption{(Color online) Extrapolation of the transition temperatures (a) $T_{\rm c1}$ and (b) $T_{\rm c2}$ with phenomenological renormalization group analysis. The broken lines in each panel are guides for the eyes. }
\label{fig:prg}
\end{figure}

\section{Entanglement spectrum and TL parameter}
\label{sec:es}

In order to reveal the nature of the intermediate critical phase and the BKT transitions, we further investigate the finite-$m$ dependence of the entanglement spectrum.
In connection to CFT for the CTM geometry, we define the entanglement Hamiltonian as 
\begin{equation}
\rho = \exp\left(- 2\pi \mathcal{H}_{\rm E} \right), 
\label{eq:HE}
\end{equation}
where $\rho$ is the reduced density matrix defined by a product of four CTMs.
Then, the conformal mapping of the boundary CFT on the upper half-plane into that for the CTM geometry [$\simeq$ the annulus with an infinitesimal inner radius] leads us to 
\begin{equation}
\mathcal{H}_{\rm E} = \frac{\pi}{\ln r/\epsilon } \left( L_0 - \frac{c}{24} \right) + {\rm const.} \, , 
\label{eq:HE_spectrum}
\end{equation}
where $L_0$ denotes the Virasoro generator of the holonomic part, $r$ and $\epsilon$ respectively correspond to the system size of the CTM and a cutoff scale~\cite{Peschel-Troung, Cardy-Tonni}.
For the $c=1$ CFT, the spectrum of $L_0$ is given by $x_{h,n}+ N $, where $x_{h,n}$ indicates possible conformal weights compatible with the boundary conditions, and $N$ is a non-negative integer corresponding to descendants~\cite{cardy_bcft}.
Then, $x_{h,n}$ is explicitly written as  
\begin{equation}
x_{h,n} = \frac{1}{2}\left( \sqrt{\frac{K}{2}}h + \frac{1}{\sqrt{2K}} n\right)^2,
\label{eq:gap}
\end{equation}
where $(h, n)$ are integer quantum numbers ($n$ corresponds to the winding number), and $K$ denotes the TL parameter representing the effect of the renormalized cosine terms in the effective Sine-Gordon theory.
For the critical phase of the six-state clock model, in particular,  $K=9$ and $K=4$ are respectively expected at $T_{\rm c1}$ and $T_{\rm c2}$, according to the analytical results for the $Z_6$ dual sine-Gordon model~\cite{Matsuo2006}.
 
For the finite-$m$ scaling analysis, the effective size of the CTM  measured from the center of the system is given by the effective correlation length, $r\sim \xi_{\rm e}$, where the free boundary condition can be assumed since the outer region of the CTM beyond $\xi_{\rm e}$ is basically decoupled from the central region of the CTMs~\cite{cho_ryu}. 
Thus, the entanglement spectrum measured from the ground state can be represented as 
\begin{equation}
\Delta E_{i} \equiv E_{i}-E_0 = \frac{\pi}{\ln \xi_{\rm e}/\epsilon }(x_{h,n} + N)
\label{eq:gap}
\end{equation}
with $i=0,1,2,3\cdots$ ($\Delta E_0=0$ corresponds to the ground state).
Note that this relation is also numerically tested for 1D critical quantum systems~\cite{Lauchli,cho_ryu}.

\begin{figure}
\includegraphics[width=7cm]{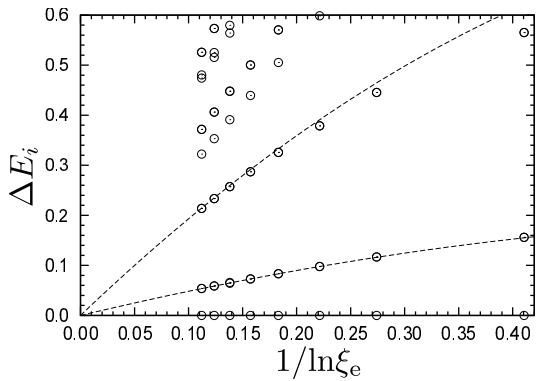}
\caption{Entanglement spectrum at $T=0.8$ as a function of $1/\ln \xi_{\rm e}$. 
The broken lines are quadratic fits with respect to $1/\ln \xi_{\rm e}$.}
\label{fig:es_vs_xi}
\end{figure}

Figure.~\ref{fig:es_vs_xi} shows the entanglement spectrum $\Delta E_i$ at $T=0.8$ as a function of $1/\ln \xi_{\rm e}$. 
The degeneracy of the spectrum is $1,2,2,1,2,\cdots$ from bottom to top, for which we can confirm the asymptotic behavior as in Eq.~(\ref{eq:gap}) in the region of small $1/\ln\xi_{\rm e}$. 
Moreover, we can read off $\Delta E_2/\Delta E_1 \simeq 4$ in Fig.~\ref{fig:es_vs_xi}, which suggests $N=0$ for consistency with Eq.(\ref{eq:gap}).
Taking account of the first excited state doubly degenerating, we may assume that the first and second excited states are respectively characterized by $x_{0,\pm 1}$ and $x_{0,\pm 2}$.

\begin{figure}
\includegraphics[width=7cm]{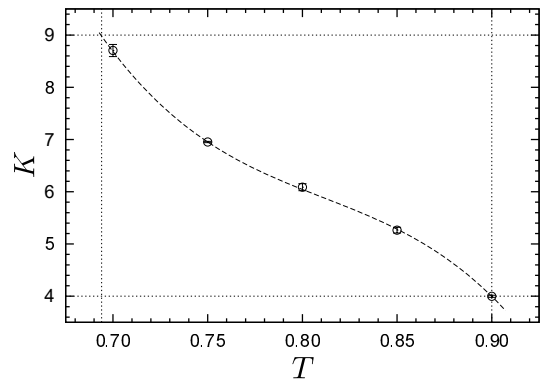}
\caption{Temperature dependence of the TL parameter $K$, which is extracted from the first excited energy of the entanglement spectrum with the quadratic fit of $1/\ln \xi_{\rm e}$.}
\label{fig:K_vs_t}
\end{figure}

In order to determine the TL parameter $K$ in the conformal weight, we perform the fitting of $\Delta E_1$ with a quadratic function of $1/\ln \xi_{\rm e}$ for $1/\ln\xi_{\rm e}<0.2$ and extract the coefficient of the leading term of $1/\ln \xi_{\rm e}$.
Assuming $x_{0,1}$ for $\Delta E_1$, we obtain the temperature dependence of $K$ in Fig. \ref{fig:K_vs_t}, which is in good agreement with the effective $Z_6$ dual sine-Gordon theory;
in particular, $K\simeq 9$ and $K\simeq 4$ can be confirmed at $T_{c1}=0.693$ and $T_{\rm c2}=0.900$, respectively.
Note that these values of the transition points are  consistent with the finite-$m$ scaling results in the previous section.

In addition, the TL parameter at the self-dual point for the $Z_6$ dual sine-Gordon model is given by $K=6$.
We evaluate $K$ for various $T$ and then find that $K=6$ is realized at $T_{\rm SD}\equiv 0.803$, which is consistent with the recent tensor network study combined with a ratio of partition functions for Klein bottles~\cite{Li2019}.
Moreover, we find that the relation $T_{\rm SD}^2 = T_{\rm c1} \cdot T_{\rm c2}$ is numerically satisfied with $T_{\rm c1}=0.693$, $T_{\rm c2}=0.900$, and $T_{\rm SD}= 0.803$ up to 3 digits, although the six-state clock model is not self-dual with respect to the Kramers-Wannier duality transformation~\cite{Savit_Rev_Mod}.
Whether this relation could be exact or an approximation in the effective field theory level is an interesting future problem.

\section{Conclusions and Discussions}
\label{sec:summary}
We have investigated the critical phenomena of the six-state clock model on a square lattice.  
We have calculated the effective correlation length, the magnetization, and  the entanglement entropy/spectrum, using the parallelized CTMRG. 
We have then performed the finite-$m$ scaling analysis, which revealed that the critical intermediate phase actually emerges, accompanying the BKT phase transitions at the phase boundaries, $T_{\rm c1}=0.694(3)$ and $T_{\rm c2}=0.908(3)$.
These transition temperatures are consistent with those of recent MC and tensor network simulations listed in Table~\ref{Table:Tc_List}. 
Also our estimation of the central charge $c$ and the exponent $\eta$ for the magnetization at $T=T_{c1}$ are basically consistent with the analytical values. 
Moreover, we have shown the low-energy behavior of the entanglement spectrum is in good agreement with the conformal dimension of the $c=1$ boundary CFT associated with the $Z_6$ dual sine-Gordon model and the resulting temperature dependence of the TL parameter is also consistent with the theoretical values~\cite{Matsuo2006, eta_for_z6} and the recent tensor network analysis~\cite{Li2019}.

However, we should note that  $\kappa=1.17(1)$ for the effective correlation length in the intermediate phase slightly deviates from  $\kappa \sim1.34$ that is expected from Eq.~(\ref{pollmann}) with $c=1$. 
Similar discrepancies are interestingly reported for one-dimensional chiral ladder, $(\kappa,c)=(1.16,0.985)$~\cite{schmoll_orus_prb_2019} and a deconfined quantum critical point, $(\kappa,c)=(1.18(3),0.99)$~\cite{huang_xiang_prb_2019}. 
A possible reason for this could be that the effective correlation length based on Eq.~(\ref{eq:correlation_length}) is not appropriate. 
As pointed out in~\cite{johnson_mccoy_pra_1973}, for example, the exact correlation length of the XYZ chain is given by integrating over the entire band of complex next-largest eigenvalues of the row-to-row transfer matrix, not by the ratio of the largest and next-largest eigenvalues. 
 In order to settle this problem of $\kappa$, further investigations on the finite-$m$ scaling analysis for the BKT transition will be needed. 
Finally, we note that  it is also an interesting problem to clarify how the phase transitions of the clock model can be connected to the critical property of the icosahedron and dodecahedron models~\cite{HU2017,pctmrg}.

\acknowledgments
This research was  partially supported by KAKENHI No. 26400387, 17H02931, and 17K14359, and by JST PRESTO No. JPMJPR1911, and by %VEGA 2/0130/15 
VEGA 2/0123/19 and APVV-16-0186. It  was also  supported by MEXT as ``Challenging Research on Post-K computer'' (Challenge of Basic Science: Exploring the Extremes through Multi-Physics Multi-Scale Simulations). 
The numerical computations were performed on the K computer provided by the RIKEN Center for Computational Science through the HPCI System Research project (Project ID:hp160262).

%\bibliography{reference}

\end{document}